\documentclass[a4paper]{article}
\usepackage[T1]{fontenc}
\usepackage[latin1]{inputenc}
\usepackage{graphics}

\makeatletter

\providecommand{\LyX}{L\kern-.1667em\lower.25em\hbox{Y}\kern-.125emX\@}

\usepackage[T1]{fontenc}
\usepackage[latin1]{inputenc}

\makeatletter

\let\SF@@footnote\footnote
\def\footnote{\ifx\protect\@typeset@protect
    \expandafter\SF@@footnote
  \else
    \expandafter\SF@gobble@opt
  \fi
}
\expandafter\def\csname SF@gobble@opt \endcsname{\@ifnextchar[
  \SF@gobble@twobracket
  \@gobble
}
\edef\SF@gobble@opt{\noexpand\protect
  \expandafter\noexpand\csname SF@gobble@opt \endcsname}
\def\SF@gobble@twobracket[#1]#2{}

\makeatletter

\makeatletter

\usepackage{setspace}
\usepackage[dvips]{graphicx}

\makeatother

\makeatother

\makeatother

\makeatother

\begin{document}

\title{HIV time hierarchy: Winning the war while, loosing all the battles.}

\author{Uri Hershberg{\scriptsize (1)},Yoram Louzoun{\scriptsize (1)},Henri Atlan{\scriptsize (2)}
and Sorin Solomon\thanks{
To whom correspondence should be sent : Prof. Sorin Solomon,Racah Institute
for Physics, Hebrew University, Jerusalem Israel. e-mail: sorin@cc.huji.ac.il.
Tel+fax: ++972-2-6585761 
}{\scriptsize (3)}.}

\maketitle
\textit{\scriptsize (1) Interdisciplinary Center for Neural Computation, Hebrew
University, Jerusalem Israel.}{\scriptsize \par{}}{\scriptsize \par}

\textit{\scriptsize (2)Human Biology Research Center Hadassah Hebrew University
Hospital, Jerusalem Israel.}{\scriptsize \par{}}{\scriptsize \par}

\textit{\scriptsize (3)Racah Institute for Physics, Hebrew University, Jerusalem
Israel.}{\scriptsize \par{}\par{}\par{}}{\scriptsize \par}

\begin{abstract}
AIDS is the pandemic of our era. A disease that scares us not only because it
is fatal but also because its insidious time course makes us all potential carriers
long before it hands us our heads in a basket. The strange three stage dynamics
of aids is also one of the major puzzles in describing the disease theoretically
\cite{1}. Aids starts, like most diseases, in a peak of virus expression\cite{2,3},
which is practically wiped out by the immune system. However it then remains
in the body at a low level of expression until later (some time years later)
when there is an outbreak of the disease which terminally cripples the immune
system causing death from various common pathogens. In this paper we show, using
a microscopic simulation, that the time course of AIDS is determined by the
interactions of the virus and the immune cells in the shape space of antigens
and that it is the virus's ability to move more rapidly in this space (it's
high mutability) that causes the time course and eventual 'victory' of the disease.
These results open the way for further experimental and therapeutic conclusions
in the ongoing battle with the HIV epidemic. 

\_\_\_\_\_\_\_\_\_\_\_\_\_\_\_\_\_\_\_\_\_\_\_\_\_\_

key words:Multi scale, Monte Carlo Simmulation, Evolution, Immunology , HIV,
Hybrid Models. 

PAC--87.22 As 
\end{abstract}

\section*{Introduction: }

The natural progression of the Human Immuno-deficiency Virus (HIV) infection
varies between individuals, however a general pattern of the progression has
been observed (Figure 1).

\begin{itemize}
\item Within weeks of infection, a short transient jump of plasma virema (virion concentration
in plasma) is seen together with a marked decrease in Immune cell counts (CD4
T Helper cells). 
\item Partial control of the disease by the immune system ensues, causing variable
periods of practically a-symptomatic clinical latency, which can last years.
During this period the Immune cell population continues to slowly decline until
the immune system is so crippled it can no longer contain the disease. 
\item A renewed outbreak of the virus which, together with constitutional symptoms
and the onslaught by opportunistic diseases, cause death \cite{1}.
\end{itemize}
\vspace{0.3cm}
{
\begin{figure}
{\par\centering \resizebox*{0.8\textwidth}{!}{\includegraphics{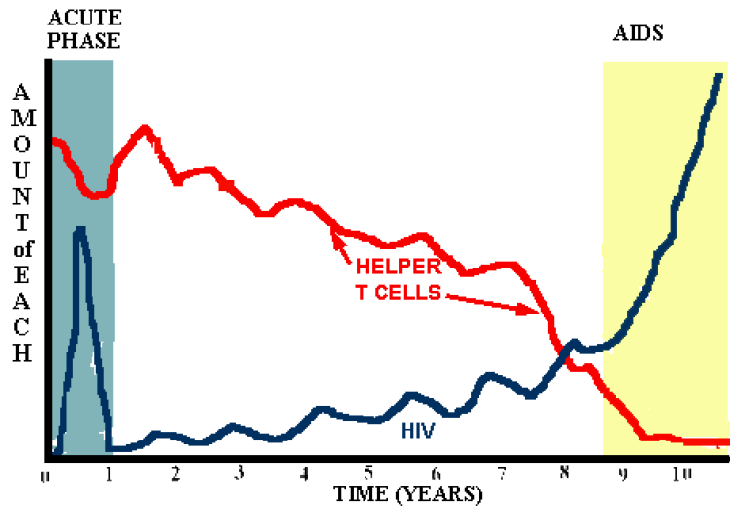}} \par}

\caption{\textbf{The typical 3 stage evolution of the HIV illness -} According to the
experimental data, the HIV infection evolution presents 3 distinct phases. During
the first months after infection, there is an acute phase with a very large
increase in the virus population and a corresponding destruction of the immune
cells. This ends with the reprise of the immune system to the invasion and the
decrease of the virus population to very low values. After the immune systems
reprise comes a long period of slowly increasing virus population and slowly
decreasing immune cells population. At some stage, the virus population rises
exponentially and the immune system collapses (the AIDS phase) resulting in
an onslaught of opportunistic diseases and death. (Addapted from picture, Copyright
Dr R.E. Hurlbert, from the Washington State University Fundamentals of Microbiology
101 course home page: http:// www.wsu.edu:8080/\textasciitilde{}hurlbert/pages/Chap16.html\#STD\_intro)\cite{4}.}
\end{figure}
\par}
\vspace{0.3cm}

The dynamics of HIV was traditionally described using simple homogeneous ODEs\cite{5}
(for a review see Perelson \cite{6}). This method was enlarged to models considering
the spatial structure mainly by Zorzenon dos Santos and Coutinho \cite{2,3}.
Such models consider the importance of the localization of the interactions
between the HIV virions and immune cells. Taking into consideration the global
features of immune response and the high mutation rate of HIV they described
the spatial and temporal interactions of infected and healthy cells in lymphoid
tissue in the body.

The methodology used in spatially extended models was limited up to now mainly
to cellular automata \cite{2,3}, or to compartmental models \cite{7}. The
main advantage of cellular automata is their capacity to emphasize the emergence
and the importance of spatial structures. For example the propagation of waves
of infection in the lymph nodes, or the creation of immune cell aggregations\cite{2,3}.
Consequently Zorzenon dos Santos and Coutinho, in contradiction with most ODE
models, succeeded to reproduce all the stages of HIV evolution, using a single
set of rules \cite[figure 13 in 2]{3}.

The present model is inspired by the observed interesting features in the spatially
extended model\cite{2,3} . Our model acts in the shape space rather than in
the physical space and exploits the dynamical implications of the HIV virus
'propagation' in the molecular shape space: The immune system's need to identify
the virus's form correctly in the shape space of possible viral and immune receptor
shapes. The main ingredients in our model are the mutations (represented by
propagation of the multiplying virions in the structure-less infinite dimensional
shape space) and the confrontation with the immune system cells (resulting in
the disappearance of the both cells and virions). Our model does not display
any geometrical patterns in either space or shape space. In this model we assume
that we can average the spatially distributed reactions (as in the ODE approach). 

The model presented in the following pages relies on the basic known facts on
the immune defense system. The viruses (antigens) have various characteristic
geometrical shapes. In order to act against them, the immune system has to 'identify'
them by producing cells which contain shapes complementary to the geometry of
(parts of) these antigens. Since the immune system does not know a priori what
is the characteristic shape of every new invading virus, the immune system generates
randomly cells with various shapes. If a cell encounters (by chance) an antigen
(virion) with complementary shape, then more cells with characteristic shapes
identical to it are produced and a mechanism is triggered for the destruction
of all the individuals (virions) belonging to this virus strain (and sharing
the same shape)\cite{8}.

Usually the destruction mechanism is quite efficient and once a virion is 'identified'
by the above random search in the shape space, its fate and the fate of all
the virions in the same strain is sealed: they are wiped out by the immune system
within days.

With HIV however, the issue is more complicated\cite{2,3}: Since the virus's
replication mechanism is relatively imprecise, as it multiplies it undergoes
a large amount of mutations/changes in shape compared to those found in other
kinds of virus \cite{9,10,11}. Based on empirical and theoretical results in
the research of HIV we propose the following scenario. The immune system cells
that are complementary to the old shape are ineffective in dealing wiht the
new mutant virus strain. The virions belonging to the strain with the new shape
can multiply with impunity until a strain of immune cells which fits the new
shape is generated by the immune system. Once the Immune system's shape generation
process succeeds to produce by chance a immune cell carrying a complementary
shape to the new virus strain and this cell encounters (by chance) a virion
belonging to the new strain, the new strain is wiped out too (with the exception
of the eventual new mutants that again can multiply freely until their new shape
is detected by the immune system). The process continues indefinitely with the
virus loosing every battle but succeeding to produce increasingly many small
populations of new shapes (figure 2)\cite{8,12,13,14,15,16} .
\begin{figure}
{\par\centering \resizebox*{0.8\textwidth}{!}{\includegraphics{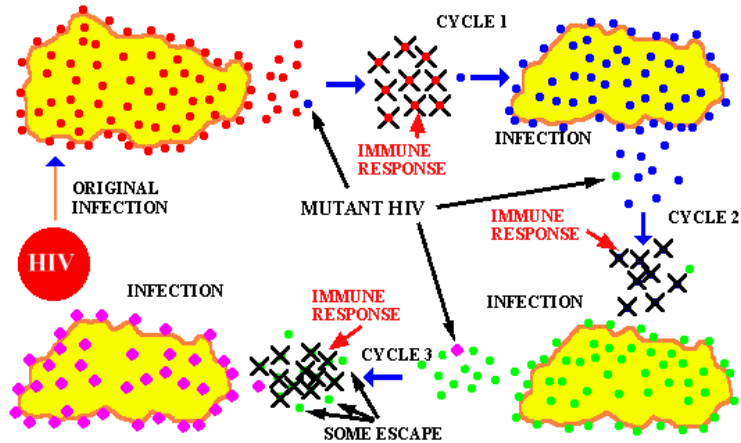}} \par}

\caption{The HIV strains in the primary infection are detected and annihilated by the
immune system. However some virions escape the immune system by mutating. While
the new population which they engender is also eventually discovered and destroyed
by the immune system, their mutating descendants will escape again. These phases
of escape and destruction enable the intermittent proliferation of HIV after
the immune systems primary response and with it the idiosyncratic time course
of HIV infection. (Picture, Copyright Dr R.E. Hurlbert, taken from the Washington
State University Fundamentals of Microbiology 101 course home page: http://
www.wsu.edu:8080/\textasciitilde{}hurlbert/pages/Chap16.html\#STD\_intro\cite{4})}
\end{figure}

The process is compounded by the additional fact that virions can kill directly
and indirectly immune cells (whether or not they are of complementary shape)\cite{17}.

The immune system continues to win every battle until the increase in production
of immune cells with shape complementary to a virus strain is overcome by the
rate at which the immune cells are destroyed by various other HIV strains. At
this point the immune system has effectively lost the war.

In the rest of the paper we provide a detailed microscopic simulation model
\cite{18,19,20,21} which supports this scenario and that fits quite well the
known phenomenological data on HIV in terms of the following basic mechanisms:

\begin{itemize}
\item The \textbf{local} (in shape space) destruction of HIV by the immune system\cite{15}. 
\item The fast mobility of HIV in shape space (high mutation rate)\cite{9,10}. 
\item The \textbf{global} destruction of immune system cells by HIV\cite{17}. 
\end{itemize}

\section{The Model}

We represent the shape space of the virus by a random lattice in which each
site (\( i \)) has a fixed number of neighbors. Neighboring sites represent
shapes that can be reached one from the other by a single base mutation of the
virus. The occupation number on each site (\( N_{Vi} \))represents the number
of virions with that shape existing in the organism. The immune system cells
that recognize that shape are also represented through an occupation number
on the same site(\( N_{Ci} \)). Note that the existence of a virus and an immune
cell on same lattice site does not imply their proximity in real space: Quite
to the contrary they might be located in very distant locations in the organism.
There is however a small probability that the virus and the corresponding cell
will meet and react in real space. Therefore each pair of virion and immune
cell located on the same lattice site has a small, but finite probability to
react (according to the rules described in detail below).

One represents the eventual mutations of the virus and the immune cells by their
rate of diffusion in the shape space (\( D_{V} \),\( D_{C} \)). More precisely
both viruses and immune cells have a certain probability for jumping between
neighboring sites.\footnote{%
The diffusion rate, and the neighborhood structure implied by the node's connectivity
can be different for the virus and for the immune cells 
}

HIV can replicate in and destroy immune cells. This is irrespective of the cell's
characteristic shape. I.e. the virus can destroy immune cells located on sites
arbitrarily far away from the site of the virus). In our model we represent
this by:

\begin{itemize}
\item A virus proliferation rate proportional to the total immune cells population
(\( C_{tot} \)). 
\item An immune cell death rate proportional to the total viral population (\( V_{tot} \)). 
\end{itemize}
We list bellow the reactions taking place in the model:

\begin{enumerate}
\item When an HIV virion and an immune cell reside on the same site the immune cell
duplicates with a rate of \( \tau _{C} \). However following realistic biological
data, we limit the multiplication rate of the immune cells (to a factor of 3
per day). 
\item When an HIV virion and an immune cell reside on the same site the virion is
destroyed with a probability rate of \( d_{V} \). 
\item Each HIV virion replicates with a rate (\( \tau _{V}C_{tot} \)) proportional
to the total number of immune cells. 
\item Each immune cell is destroyed with a probability rate (\( d_{C}V_{tot} \))
proportional to the total number of virions. 
\item New immune cells, with various shapes are created continuously. We represent
this by a probability rate (\( \lambda  \)) for an immune cell to appear on
a random lattice site. 
\item Both the immune cells and the HIV virus diffuse slowly in the shape space with
rates \( D_{V} \) and \( D_{C} \) . 
\end{enumerate}

\section{Results}

The reactions listed in the previous section lead to the following scenario
for the evolution of the HIV infection. The virus enters the body in a high
concentration limited to a restricted number of strains. As long as there is
no immune cell in the site corresponding to a certain strain, the virions located
in this strain site proliferate exponentially. The rise in virus concentration
is accompanied by a destruction of random T cells hosting the virus (according
to item 4). These T cells can be located anywhere in the lattice. Eventually
one of the immune cells generated by the immune system will fall by chance on
the site of this strain (according to item 5) . This immune cell will proliferate
very fast, since, according to item 1, the proliferation rate increases with
the local viral concentration. The resulting high local immune cells concentration
will destroy all the virions of this strain (item 2). As a result, after a short
period (1-2 month) all the initial strains will be discovered by the immune
system and will be destroyed ending the acute phase of the disease. In the absence
of virus diffusion in shape space (i.e. mutations), this would stop the disease
(Figure 3).

\vspace{0.3cm}
{
\begin{figure}
{\par\centering \resizebox*{0.8\textwidth}{!}{\includegraphics{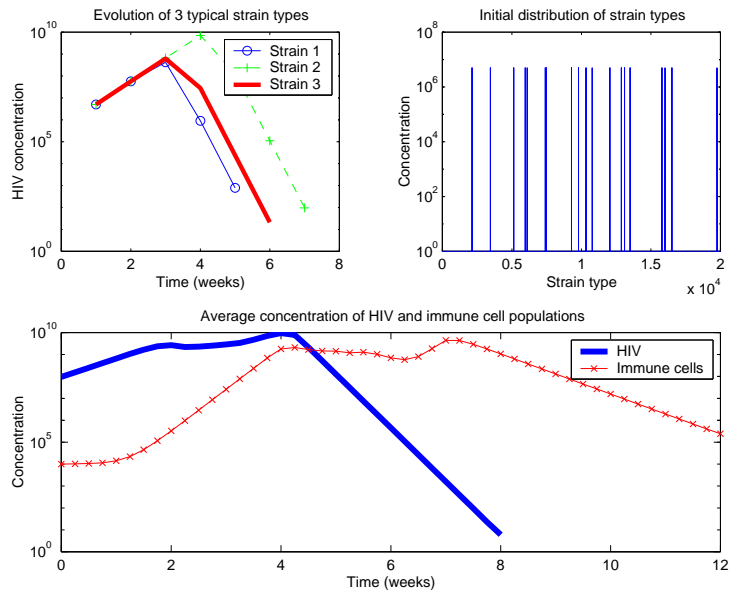}} \par}

\caption{\textbf{The acute phase analyzed by strains-} The simulation of the acute phase
provides an explanation to the very large peak in the virus population: In HIV
infection as opposed to an infection by a single virus strain, the organism
has to discover a multitude of strains. The height of the virus population peak
is usually dominated not by the average strain population but by the population
of the strain which is discovered last by the immune system (and has the longest
time to exponentiate). In the simulation the acute phase starts with a constant
concentration of HIV distributed between 20 strains (upper right drawing). Each
one of these strains activates an immune response and is destroyed (upper left
drawing). The average over all strains is the observed average HIV and immune
cell concentration (Lower drawing).}
\end{figure}
\par}
\vspace{0.3cm}

It is the diffusion rate of the virus in shape space (item 6) which is responsible
for the continuation of the infection. Before all the initial strains are destroyed,
some of the virions have a chance to mutate and escape to lattice points containing
no immune cells.

Each of these new lattice points in the shape space will contain a lower concentration
of virus than the original infection. Since the probability for the discovery
of a virus strain by the immune cells is proportional to the virus strain concentration
(according to item 1) the time it will take for the immune system to discover,
and destroy these new strains (item 2) will be longer than in the acute phase.
By the time these new strains are destroyed some virions from these strains
will have diffused to neighboring sites (undergo mutations), and constitute
the germs for a new generation of emerging strains. These strains in turn will
have the fate of their predecessors in the previous generation. One sees now
that one can describe the long-term evolution of the HIV infection as an iterative
process. More precisely the long-term evolution will consist of a chain of small
infections, each of which is easily defeated by the organism. However after
each such infection the number of strains will grow. Therefore even though the
number of virions in each strain is always kept under control by the organism,
the total amount of virions will stochastically slowly increase. The increase
is stochastic, since it depends on the random time it takes the immune system
to discover and destroy each new strain. As the number of virions increases,
so does the death rate of the immune cells (item 4) . At a certain stage the
death rate will be high enough to impair the capacity of the immune system to
react locally to new strains. This constitutes the last stage of the disease
(Figure 4).

\vspace{0.3cm}
{
\begin{figure}
{\par\centering \resizebox*{0.8\textwidth}{!}{\includegraphics{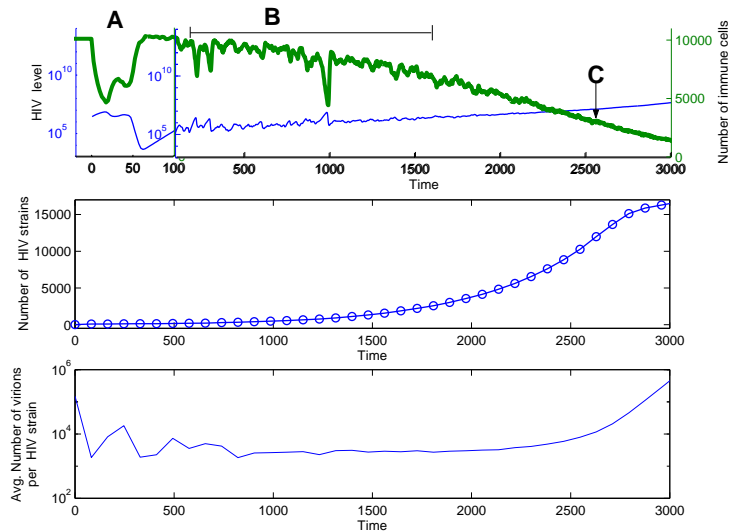}} \par}

\caption{\textbf{Typical HIV evolution in our simulation-} In the top part of the figure
we see that with our simulation we capture the dynamics found in the empirical
data. By comparing with the Fig. 1, one sees that our model reproduces correctly
the 3 stages of the disease. The acute phase (A), a phase of chronic latency
(B) and finally a renewed outbreak of HIV coupled with the destruction of the
immune system (C). Note that the first 100 days are plotted at a different scale.
In the two other figures we see, over the progression of the disease, the number
of different HIV strains (middle) and the average concentration of viroins per
strain (bottom). Together these two graphs strengthen our explanation that the
time course of HIV is based on the contradiction in local and globals concentration
of HIV - Locally the levels of HIV are kept by the immune system at a minimum
far below the level found in the primary infection. Globally, through the rise
in the number of strain types, the virus is spreading stochastically through
shape space and increasing its numbers by avoiding the local immune attacks.}
\end{figure}
 \par}
\vspace{0.3cm}

This typical scenario can vary from person to person:

\begin{enumerate}
\item The typical case observed in most hosts is a disease composed of 3 stages (Figure
4). The first acute stage is due to the fast proliferation of the original strains
before the appropriate immune response is set. This stage ends when the appropriate
immune response to each and every original strains is completed. This stage
takes approximately \( \lambda * \)ln(Number of original strains) days (see
below). This gives an expected number of virions at the peak, which is much
higher than in a usual disease (Appendix A) . The latent stage is the stage
in which the number of virion strains is limited, and the number of virions
in each strain is low. This stage will last as long as the number of strains
is much smaller than \( {\tau _{C}\over d_{C}} \) (Appendix B). When the number
of \textbf{strains} grows above \( {\tau _{C}\over d_{C}} \) the last stage
of the disease occurs. This is the regime in which the local (in shape space)
activation of the immune system by the local virus strain (item 2) becomes lower
than the global destruction rate of immune cells by the virions (item 4), and
the immune system fails to destroy \textbf{locally} (item 4) the existing strains.
At this stage many old strains that were kept under control during the previous
stages can reappear. 
\item If the efficency of immune reaction to HI (items 1 and 2) is high enough, the
immune system will manage to defeat the new virus strains fast enough. Thus
the average number of strains will stay constant or slowly decrease. In this
case the total number of virions will vary around a fixed number, or slowly
decrease to 0. This fate can be the one of the long time carriers8\cite{22,23}.
In reality, the total number of virions never decreases to zero, since there
are other sources (macrophages, neuronal cells, etc.) that contribute a small
number of new virions \cite{22,24}(Figure 5). 
\item When, on the other hand, the efficency of immune reaction is low a large number
of new strains will be created before the immune cells will finish destroying
the virions from the strains of the first infection. In this case the acute
phase will directly lead to the death of the host (Figure 6)\cite{22,25}. 
\begin{figure}
{\par\centering \resizebox*{0.8\textwidth}{!}{\includegraphics{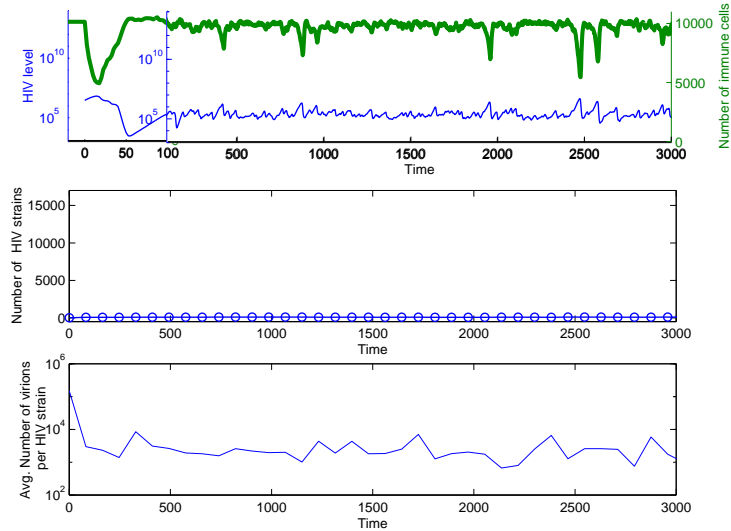}} \par}

\caption{Merely by raising the efficiency of immune cell reaction to HIV at the different
sites we can bring about a time course suggesting the progression of HIV in
long-term latents \cite{23}.}
\end{figure}

\end{enumerate}
The results of our simulations show that although each new strain that emerges
is destroyed within a few weeks (like any ordinary disease), the long term evolution
of the infection takes years. The time scale that determines the long-term evolution
scale is the diffusion (mutation) rate of the virus. The number of new strains
growes exponentially with time , but the unit of time in this exponential is
the time it takes for a new strain to establish itself i.e. weeks.

One might ask how the scale hierarchy between the cellular interactions (hours)
and the evolution of the infection (years) emerges in this model. The answer
is the following. The single strain lifetime is determined by an exponential
rate proportional to the local interaction rate (hours). This exponent rises
to macroscopic virus concentration within weeks. Only when the value of this
exponent is high enough does the long-term mechanism of virus mutation become
operative. Thus the time unit for one interaction in the shape space is the
time needed for a single strain to establish itself. The time scale of the entire
disease is the time necessary for evolving a macroscopic number of strains.
Therefore we expect that the time scale of the entire disease will relate to
the time scale of new strain creation as the time scale of the strains relate
to the individual virion division time. The actual numbers are indeed (13 years/
2 weeks) = (2 weeks/hours).
\begin{figure}
{\par\centering \resizebox*{0.8\textwidth}{!}{\includegraphics{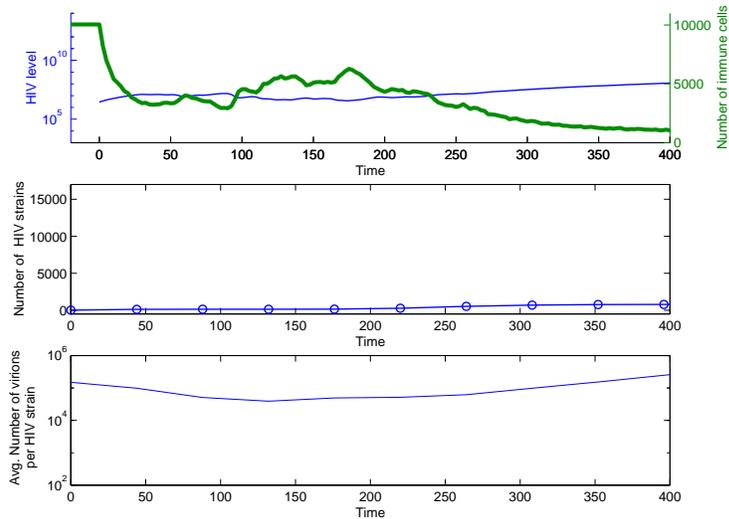}} \par}

\caption{By decreasing the efficiency of immune cells we arrive at a scenario where
already in the first acute phase we witness a full-fledged outbreak of AIDS.
This is reminiscent of empirical evidence to the fact that quick progressors
of HIV have a less flexible immune system and thus succumb to the disease with
less strains of the virus in their bloodstream \cite{25}. (Note that since
death is so rapid the time course is only measured untill the colapse of the
immune system around day 400).}
\end{figure}

\section{Conclusions}

The main success of the present model is the \textbf{natural} emergence of a
hierarchy of very different dynamical time scales. The very long-term decrease
in immune (CD4+ T) cells count cannot be explained by a simple dynamic system.
The transition from the microscopic time scales (hours) to the macroscopic time
scales (years) requires a profound explanation. A simple dynamical system would
require extreme fine tuning of its parameters in order to achieve such a transition\cite{2,3}.
We propose a mechanism that can bridge between the microscopic and macroscopic
time scales, that does not need fine tuning. The transition can be expressed
best by representing the evolution of the system in the shape space (the strain
of the virus), rather than the real space (the location of the virus in the
organism). The relavent unit of time step for the operations in shape space
is the time it takes for a strain to reach a macroscopic concentration and therefore
have a significant probability to generate a mutant. This time step is of the
order of weeks and not of hours. The evolution of the disease takes a few hundred
time steps. i.e a few hundreds of weeks.

The mechanisms operating at the short times and at the long times are completely
different. At the microscopic level the mechanism is the recognition and the
destruction of the virions by the immune cells. The short time scale evolution
of the disease is similar to any other disease. The long scale evolution of
HIV infection is based on the competition between localized (in shape space)
processes and global processes. To be precise the evolution is due to the spread
of the virus strains across the shape space. In the initial acute phase most
of the viral load is distributed between a small number of localized virus strains.
At the last stages of the disease the viral load is distributed between a large
variety of many strains. The immune cells manage to destroy locally every particular
virus strain within a couple of weeks from its emergence. However as the number
of strains at each given moment increases, the virus succeeds to destroy an
increasing amount of immune cells. Thus locally the virus looses every battle.
Yet in the end the shape space is filled with a multitude of small but numerous
strain populations. At that point the virus wins the war by killing more immune
cells than it activates.

In short: During the latent stage every virus strain looses every battle since
it is activating the cells that can destroy it. Transition to AIDS occurs when
the combined contemporary virus population wins the war by killing in total
more immune cells than the sum of cells it activates.

\section*{Acknowledgment}

We acknowledge Professor Zorzenon dos Santos for sharing with us the content
of reference 3 before publication.

\section*{Appendix A - Acute phase}

Imagine one has a lattice containing V nodes out of which N are occupied by
the strains that constitutes the initial infection. We assume that the population
of each initial infection strain is large enough that if the immune system generates
randomly an immune cell in this site (A cell with a complementary shape to that
virus strain) then that cell will discover this strain with probability 1. The
probability rate for generating in a given lattice site an immune cell is \( \lambda  \).
Therefore the probability rate for discovering one of the strains is \( N\lambda  \).
Thus the average time it takes the immune system to detect the first strain
in the initial infection is \( {1\over N\lambda } \). Once this strain was
discovered the probability rate to discover one of the remaining (N-1) strains
is \( (N-1)\lambda  \). This means that the time it takes to discover a second
strain is \( {1\over (N-1)\lambda } \) and so on. Thus the average time it
will take for the immune system to discover the N strains is: \( {1\over lambda}\Sigma _{1}^{N}{1\over i} \)
which is approximately \( {ln(N)\lambda } \). As long as the virus strain is
undetected by the immune system it can proliferate freely. Therefore the strain
discovered last had the most time to proliferate. The factor by which it grew
more than a single strain infection is : \begin{equation}
e^{{ln(N)-1\over \lambda }*\tau _{V}*C_{total}}\cong N^{\tau _{V}*C_{total}\over \lambda }
\end{equation}

\section*{Appendix B - Transition to AIDS}

In this appendix we estimate the conditions for the final collapse of the immune
system, when it fails to react appropriately to local virus strains. The dynamics
of the population of the immune cells in a given site in the shape space is
dominated by 2 parallel processes:

\begin{itemize}
\item Proliferation due to the activation of the immune cells by the interaction with
the local (in shape space) virus strain (\( \tau _{C}V_{i} \))(item 1). 
\item Random global destruction by arbitrarily shaped virus strains (\( d_{C}V_{tot} \))
(item 4). 
\end{itemize}
An immune strain can increase its population if its proliferation rate is higher
than its destruction rate. \( \tau _{C}V_{i}>d_{C}V_{tot} \) . In other words
we need the ratio between local virus concentration and global virus concentration
to be :\( {V_{tot}\over V_{i}}<{\tau _{C}\over d_{C}} \). If we have N virus
strains proliferating in the system, we can assume that their population are
of the same order of magnitude, and estimate \( {V_{tot}\over V_{i}}=N_{strains} \).
Thus in order for an immune strain to be able to rise its concentration and
react appropriately against the corresponding virus strain One needs the number
of virus strains not to exceed : \( N_{strains}={\tau _{C}\over d_{C}} \).
If one assumes that the virus population is not equally divided between strains
the inequality is even more stringent.

\section*{Appendix C - Discretization vs Continuous Differential Equations}

The evolution of a dynamical system can be expressed by 2 types of models:

\begin{enumerate}
\item Ordinary Differential Equations (ODE) that simulate the evolution of the average
population under the assumption of spatial homogeneity. 
\item Microscopic simulation (MS) models, that compute each reaction separately. 
\end{enumerate}
The ODEs have the advantage of being cheap in CPU time. They enable us to simulate
precisely the system when its concentration is very high, but fail to describe
the stochastic aspects of the system. The MS takes into account the stochastic
effects and describes precisely the discrete aspects of the agents and of the
strains. However MS is very inefficient if the number of agents and the probability
for reaction are high.

In the present model most of the sites are basically empty. However there is
a relatively small number of sites occupied by a macroscopic number of virions
and immune cells. This special situation invalidates both the possibility to
use continuity assumption (ODE), and discrete operations (MS).

We solved this problem by using a hybrid model. This model computes the probability
for interaction between every 2 agents at every site on the lattice in a given
time interval. If this probability is higher than a threshold (30) then the
number of agents created or destroyed is computed in a deterministic way using
an ODE formalism \textbf{for this site}. If on the other hand the probability
for a reaction is lower than the threshold the number of new agents created
or destroyed is computed in a discrete stochastic way.

This is a particular application of the hybrid models we developed \cite{18,19,20,21}.

\end{document}